# XPS studies of nitrogen doping niobium used for accelerator applications

ZiqinYang[a,*], Xiangyang Lu[b], Weiwei Tan[b], Jifei Zhao[b], Deyu Yang[b]
Yujia Yang[b], Yuan He[a], Kui Zhou[c]

[a]Institute of Modern Physics, Chinese Academy of Science, Lanzhou 730000, China
[b]State Key Laboratory of Nuclear Physics and Technology, Peking University, Beijing 100871, China
[c]Institute of Applied Electronics, Chinese Academy of Engineering Physics, Mianyang 621900, China

**Abstract**

Nitrogen doping study on niobium (Nb) samples used for the fabrication of superconducting radio frequency (SRF) cavities was carried out. The samples' surface treatment was attempted to replicate that of the Nb SRF cavities, which includes heavy electropolishing (EP), nitrogen doping and the subsequent EP with different amounts of material removal. The surface chemical composition of Nb samples with different post treatments has been studied by XPS. The chemical composition of Nb, O, C and N was presented before and after Gas Cluster Ion Beam (GCIB) etching. No signals of poorly superconducting nitrides $NbN_x$ was found on the surface of any doped Nb sample with the 2/6 recipe before GCIB etching. However, in the depth range greater than 30nm, the content of N element is below the XPS detection precision scope even for the Nb sample directly after nitrogen doping treatment with the 2/6 recipe.

*Keywords:* Niobium, Nitrogen doping, X-ray photoelectron spectroscopy, Surface composition, Nitride

## 1. Introduction

The Nb SRF cavities has been widely used in many particle accelerator projects, which cover the application range from high energy colliders [1-3] to light sources [4-6] and small scale industrial applications[7,8]. The unloaded quality factor Q is one of the most important factors of SRF cavities, which is inversely proportional to the cavities' surface resistance. High quality factor can efficiently decrease the cryogenic load of SRF cavities. The post treatment has an important influence on the cavities' radio frequency (RF) performance. Niobium used for cavity production undergoes several different treatments before it eventually becomes a SRF cavity ready for test. The typical procedures not only include rolling, deep drawing, electron beam welding and other mechanical treatments, but also include chemical polishing, high temperature degassing, high pressure rinsing, $120^0$C baking and other post treatments. The magnetic field in the superconductor decays exponentially. Thus the subtle material details and chemical composition of the niobium surface largely influence the surface resistance of the SRF cavities below 2K. XPS analysis of the surface composition of niobium used for the fabrication of SRF cavities after procedures commonly employed in the preparation of Nb SRF cavities have been reported [9-10] and obtained better understanding of the influence of different post treatments on the Nb cavities' RF performance.

Nitrogen doping is a new surface treatment discovered by A. Grassellino [11], which can systematically improve the quality factor of Nb SRF cavities up to a factor of about 2-4 compared to the standard surface treatment procedures. Presently, nitrogen doping treatment is being transferred from the prototyping and R&D stage to the production stage [12]. Fundamental understanding of the nitrogen doping mechanism is being carried out extensively [13-15], but yet remains unclear. The Nb SRF cavities directly after nitrogen doping treatment showed quality factor in the range of $10^7$ at 2K, which is far below the routinely achieved Q values of Nb cavities with the standard surface treatment procedures at the same frequency and temperature. This has always been thought to be caused by the formation of unwanted poorly superconducting nitrides [16].

The aim of the present study is to investigate the elemental composition and chemical states both on the surfaces and in the penetration depth of the niobium samples with different treatments. The X-ray Photoelectron Spectroscopy is suitable to provide this information. For this reason XPS studies of both doped and un-doped samples before and after GCIB etching has been carried out, with particular attention paid to the elemental composition and chemical states of nitrogen both on the surface and in the penetration depth range.

## 2. Experimental

### 2.1 Sample preparation

All the Nb samples were from the same high purity (RRR~300) niobium sheet. Instead of wire electro-discharge machining (EDM), the niobium samples were manually processed to avoid heat generation. The impurity contents are listed in Table 1. After been polished smoothly using 180-grit sandpaper, 300-grit sandpaper and 1200-grit sandpaper, the niobium strips were etched by EP (HF:$H_2SO_4$ =1:9), with the material removal of about 150μm, to remove the mechanical damage layers and surface contaminations introduced during handling or exposing to the air.

**Table 1**. Contents of the main impurities in Nb samples (ppm wt).

| C | N | H | O | Ta | Fe | Si | W | Ni | Mo | Ti |
|---|---|---|---|----|----|----|----|----|----|----|
| 5 | 8 | 2 | 8 | <100 | 5 | 10 | <10 | <5 | <10 | <5 |

Before the heavy EP removal of 150μm, the niobium strips were cleaned in the following steps:
- Degreased in a solution of ultrapure water and detergent with ultrasonic agitation for 10 minutes.
- Rinsed in the ultrapure water with ultrasonic agitation for 5 minutes.
- Rinsed in the anhydrous alcohol with ultrasonic agitation for 5 minutes.
- Dried by nitrogen blowing to the surface of the niobium strips.

After the heavy EP removal, the niobium strips were rinsed with ultrapure water and anhydrous alcohol again and then were kept in the vacuum environment for nitrogen doping experiment.

Vertical test results of N-doped SRF cavities with varying doping parameters showed that the 2/6 recipe proposed by Fermi National Accelerator Laboratory (FNAL) is the best nitrogen doping recipe [12, 17]. The study in this paper is focused on this recipe. The nitrogen doping treatment of the niobium samples was carried out in a furnace using a high purity quartz tube, taking the advantage of high purity quartz's good thermal and chemical stability at high temperature. The furnace's background pressure is better than $5\times10^{-5}$ Pa, which can meet the requirement of nitrogen doping experiment. After three hours of degassing at $800^0$C in high vacuum, the samples were kept at $800^0$C for two minutes in a nitrogen atmosphere of about 40mTorr, followed by another six minutes at the same temperature in high vacuum. After six minutes of thermal diffusion, the furnace was turned off to cool down naturally. Some samples with only $800^0$C heat treatment were used for comparison. The temperature and pressure of both nitrogen doping treatment and $800^0$C heat treatment is shown in Fig.1.

After nitrogen doping treatment and $800^0$C heat treatment, the niobium samples were followed by subsequent EP with different amounts of material removal to study the effects of different contents of nitrogen impurity on the surface chemical composition of niobium.

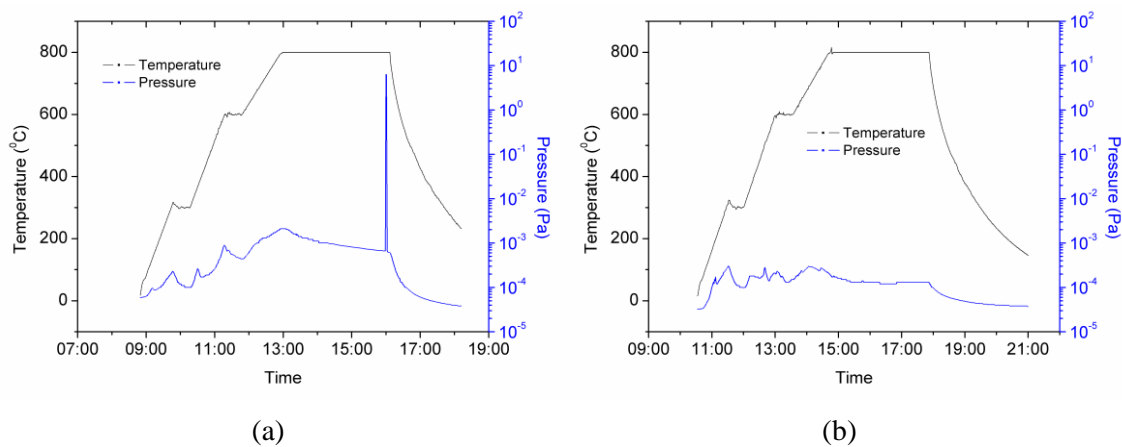

(a) (b)

**Fig. 1.** The temperature and pressure of samples with (a) nitrogen doping treatment and (b) $800^0$C heat treatment.

*2.2 XPS measurements before GCIB etching*

The elemental composition and chemical states on the sample surface was analyzed by using a Thermo Scientific ESCALab250Xi Multifunctional Photoelectron Spectrometer. The photoelectrons are excited by using a monochromatized source that produces Al Kα (hν=1486.6eV) radiation and the power is about 200W. The surface area of analysis of the sample is about 500μm×500μm. A base pressure of about $3\times10^{-10}$ mbar is obtained in the analysis chamber. Both doped and un-doped samples from two nitrogen doping experiments were chosen for the XPS studies. The samples analyzed before and after GCIB etching has been summarized in Table 2.

**Table 2**. Samples analyzed before and after GCIB etching.

| Before GCIB etching | After GCIB etching |
| --- | --- |
| ND-1st-0μm sample[1] | ND-2nd-0μm sample |
| ND-2nd-0μm sample | ND-2nd-7μm sample |
| ND-2nd-1μm sample[2] | ND-2nd-21μm sample |
| ND-2nd-7μm sample | HT-2nd-20μm sample |
| ND-2nd-13μm sample | noNDnoHT sample |
| ND-2nd-21μm sample | |
| HT-2nd-0μm sample[3] | |
| HT-2nd-20μm sample | |

| | noNDnoHT sample[4] |
|---|---|

[1] N-doped sample from experiment 1 with 0μm subsequent material removal

[2] N-doped sample from experiment 2 with 1μm subsequent material removal

[3] $800^0C$ heat treated sample from experiment 2 with 0μm subsequent material removal

[4] The sample after EP 150μm without neither nitrogen doping treatment nor $800^0C$ heat treatment.

A lower resolution survey over the wide energy window, coving the range of 1350eV with the bin size of 1 eV, was used to get information about elements presentation on the sample. High resolution scans around peaks corresponding to the elements of interest were then performed with the bin size of 0.05 eV to obtain the fine structure of the peaks, which contain the information about the chemical environment.

*2.3 XPS measurements after GCIB etching*

RF electromagnetic field is mainly distributed in the penetration depth range of the SRF cavities inner surface. Depth profiles of impurity elements in doped samples by SIMS [18-19] revealed that the concentration of C, N, O decreases with incremental EP. This may result in a different chemical composition in the penetration depth than that on the surface before GCIB etching. To obtain the elemental composition and chemical states in the penetration depth, it is necessary to etch a certain thickness on the surface of each sample, and then apply XPS study to the new surface.

Conventional Ar ions was often used for XPS depth analysis, but the energy of a single Ar ion is too high, it is easy to damage the surface of the sample, thereby changing the surface composition [20]. Due to the inherent properties of high sputtering yields and low energy bombardment effects [21], GCIB etching was applied to the investigation of elemental composition and chemical states in the penetration depth in this paper, with the expectation that the energy per atom of the GCIB clusters could be suppressed to a level lower than the chemical bond energies of the samples and not cause dissociation, chemical reaction or ion deposition within the samples. Argon was used in GCIB treatment because of its chemically inert and inexpensive. Argon clusters bound together with Van der Waals forces have average cluster size of about 500 atoms and the average cluster charge is +1. Three times of etchings were performed on each sample surface. The average cluster energy of each etching is 5keV, 10keV, 15keV, respectively. For each time of GCIB etching, the etching time is 30 minutes. The corresponding etching depth of 5keV Ar cluster for 30 minutes is about 30nm. After each time of etching, a lower resolution survey and higher resolution scans were performed to the new surface.

## 3. Results

*3.1 On the surface*

The XPS survey spectra of the ND-$2^{nd}$-0μm sample, the ND-$2^{nd}$-7μm sample, the HT-$2^{nd}$-20μm sample and the noNDnoHT sample are shown in Fig. 2 for reference. There are visible peaks for oxygen O 1s, niobium Nb 3d, Nb 3p, Nb 3s, Nb 4s and carbon C1s in the survey spectra of all samples. Very weak nitrogen signals N 1s was detected in the survey spectra of nitrogen doping samples, but there is no peak of nitrogen N 1s noticeable in the survey spectra of un-doped samples. Also, a trace amount of fluorine F 1s was detected in the survey spectra of samples after EP removal, while there is no trace of fluorine F 1s in the surface layer of samples without EP removal.

The samples were prepared from the same niobium sheet and have been cleaned in the same steps before and after EP. All the samples were kept in the same vacuum environment before the XPS measurements and all the XPS measurements were performed at the same time. So even though contamination with carbon is commonly known, it is reasonably to be stressed that the contamination coming from the hydrocarbon compounds were the same in each case. Therefore, the hydrocarbon C1s line at 284.8eV from adventitious carbon was used for energy referencing.

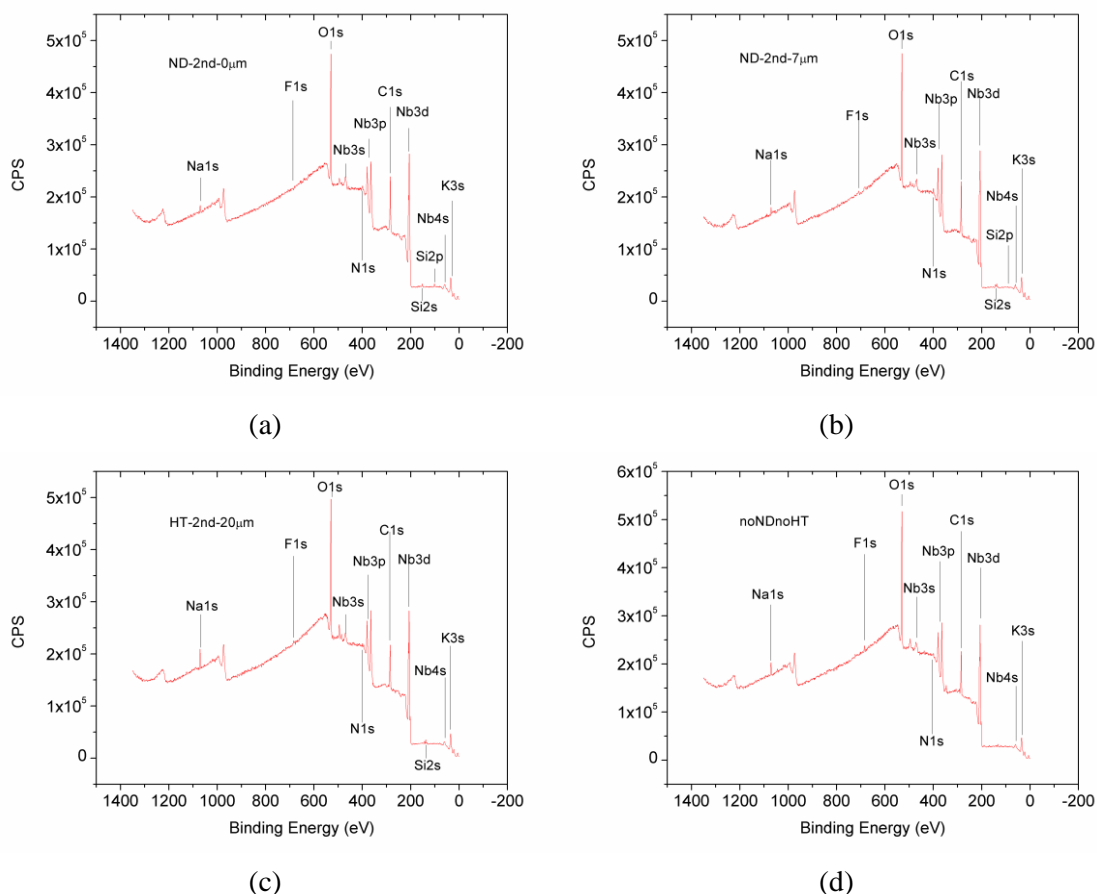

(a)      (b)

(c)      (d)

**Fig. 2.** XPS survey spectra of the ND-2$^{nd}$-0μm sample (a), the ND-2$^{nd}$-7μm sample (b), the HT-2$^{nd}$-20μm sample (c) and the noNDnoHT sample (d).

High resolution XPS spectra for niobium Nb 3d was carried out within 196-216 eV. High resolution XPS spectra for oxygen O 1s was carried out within 523-543 eV. High resolution XPS spectra for carbon C 1s was carried out within 278-298 eV. High resolution XPS spectra for nitrogen N 1s was carried out within 392-412 eV.

**Table 3**. XPS surface analysis results of samples with different treatments.

| Sample | Core level | Area (CPS. eV) | SF | Area of SF | Content (at %) |
|---|---|---|---|---|---|
| ND-2$^{nd}$-0μm | Nb 3d | 210017.32 | 9.413 | 22311.41 | 16.62 |
| | C 1s | 58411.28 | 1 | 58411.28 | 43.51 |

|  | O 1s | 141133.75 | 2.881 | 48987.76 | 36.49 |
|  | N 1s | 5273.50 | 1.676 | 3146.48 | 2.34 |
|  | F 1s | 0 | 4.118 | 0 | 0 |
| ND-2$^{nd}$-7μm | Nb 3d | 214736.17 | 9.413 | 22812.72 | 15.78 |
|  | C 1s | 63343.35 | 1 | 63343.35 | 43.82 |
|  | O 1s | 145313.97 | 2.881 | 50438.73 | 34.89 |
|  | N 1s | 4872.68 | 1.676 | 2907.33 | 2.01 |
|  | F 1s | 1555.41 | 4.118 | 377.71 | 0.26 |
| HT-2$^{nd}$-20μm | Nb 3d | 212585.96 | 9.413 | 22584.29 | 17.16 |
|  | C 1s | 49461.00 | 1 | 49461.00 | 37.59 |
|  | O 1s | 155264.95 | 2.881 | 53892.73 | 40.96 |
|  | N 1s | 3541.86 | 1.676 | 2113.28 | 1.61 |
|  | F 1s | 2069.70 | 4.118 | 502.60 | 0.38 |
| noNDnoHT | Nb 3d | 198815.27 | 9.413 | 21121.35 | 15.37 |
|  | C 1s | 52153.22 | 1 | 52153.22 | 37.95 |
|  | O 1s | 162395.46 | 2.881 | 56367.74 | 41.02 |
|  | N 1s | 3074.25 | 1.676 | 1834.28 | 1.33 |
|  | F 1s | 6193.20 | 4.118 | 1503.93 | 1.09 |

The atomic concentration of the elements present on the surfaces of samples with different treatments was determined by using the relative sensitivity factor [22]. The analysis of XPS results concerning core levels of Nb 3d, O 1s, C 1s, F 1s, and N 1s is presented in Table 3. A higher percentage of nitrogen of about 2% exists on the surfaces of doped samples, resulting from the diffused nitrogen during the nitrogen doping treatment. There is also trace of fluorine on the surfaces of niobium samples after EP removal, which is introduced in the process of EP. The atomic concentration on the surfaces of other test samples is similar to these listed in Table 3.

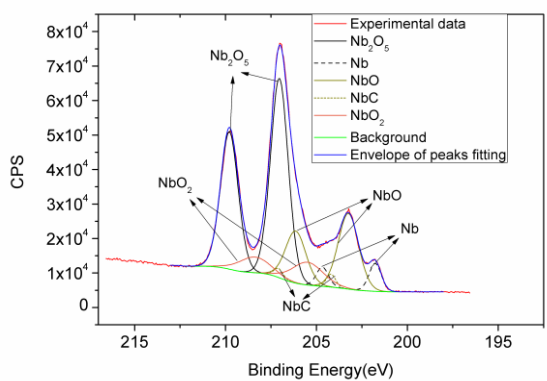 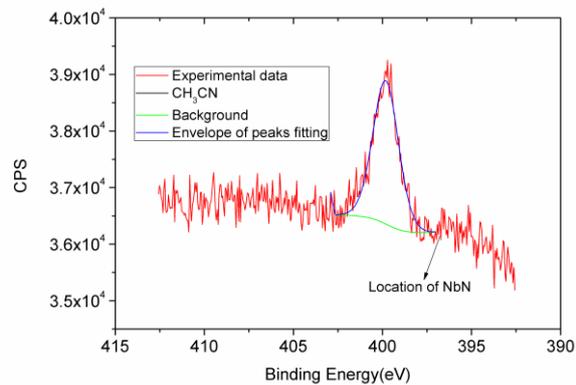

(a)                                      (b)

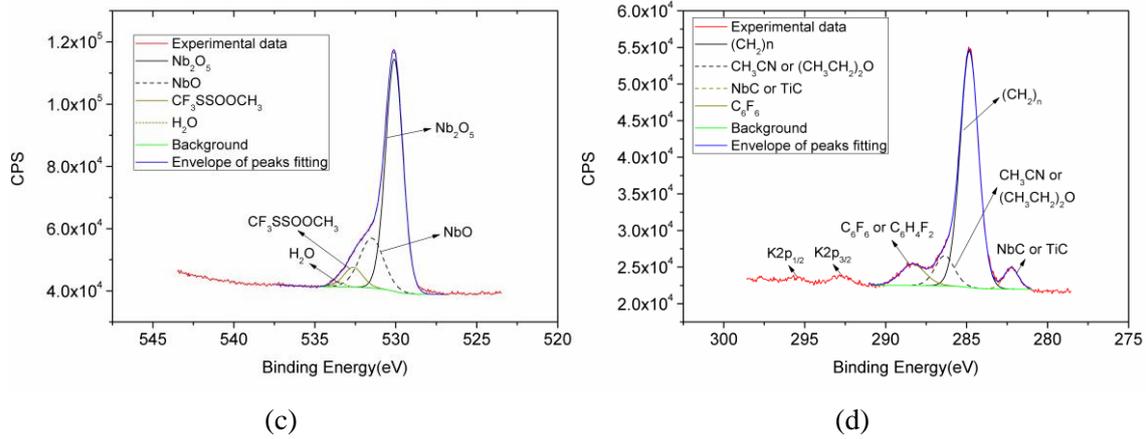

(c)             (d)

**Fig. 3.** High resolution XPS spectra and peak fitting in the narrow range of the binding energy for Nb 3d (a), N 1s (b), O 1s (c) and C 1s (d) on the surface of ND-2$^{nd}$-0μm sample.

Much more detailed analyses of the composition and chemical states of elements on the sample surfaces were carried out from the narrow scan spectra with higher energy resolution of 0.05 eV. The XPS has been traditionally used to analyze the chemical states of atoms on the surface of solid bodies [23-24]. High resolution XPS spectra in the narrow range of the binding energy for Nb 3d, N 1s, O 1s and C 1s on the surface of ND-2$^{nd}$-0μm sample are presented in Fig. 3. Also shown in Fig. 3 is the peak fitting of the high resolution XPS spectra. The background type was selected as smart. An area constraint was applied to the doublet of Nb 3$d_{5/2}$ and Nb 3$d_{3/2}$ bands. The L/G ratio was restricted between 10 and 40. The FWHM of all peaks was basically smaller than 2.5 eV. The standard values of electron binding energy of elements in different chemical environments from NIST X-ray Photoelectron Spectroscopy Database [25] were taken as reference. The detailed composition and chemical states of elements on the surface of ND-2$^{nd}$-0μm sample is presented in Table 4.

**Table 4**. Composition and chemical states of elements on the surface of ND-2$^{nd}$-0μm sample.

| Elements | | Composition and chemical states | | | | |
|---|---|---|---|---|---|---|
| Nb 3$d_{5/2}$ | | $Nb_2O_5$ | Nb | NbO | $NbO_2$ | NbC |
| | BE (eV) | 207.04 | 201.77 | 203.24 | 205.66 | 204.28 |
| | FWHM(eV) | 1.14 | 0.79 | 1.32 | 2.02 | 0.60 |
| | At % | 56.93 | 5.66 | 24.09 | 8.76 | 4.56 |
| O 1s | | $Nb_2O_5$ | NbO | $CF_3SSOOCH_3$ | $H_2O$ | |
| | BE (eV) | 530.08 | 531.11 | 532.48 | 533.53 | |
| | FWHM(eV) | 1.18 | 2.51 | 1.53 | 0.81 | |
| | At % | 59.58 | 34.28 | 5.56 | 0.59 | |
| C 1s | | Hydro-carbon | $CH_3CN$/$(CH_3CH_2)_2O$ | NbC/TiC | $C_6F_6$/$C_6H_4F_2$ | |
| | BE (eV) | 284.83 | 286.36 | 282.25 | 288.33 | |
| | FWHM(eV) | 1.33 | 1.40 | 1.06 | 1.78 | |
| | At % | 75.51 | 10.04 | 5.35 | 9.09 | |

From Fig. 3 and Table 4, it can be seen that niobium in the surface layer formed after nitrogen doping treatment with the 2/6 recipe without subsequent EP removal mainly occurs in 56.93 (at %) as $Nb_2O_5$ for BE = 207.04 eV (FWHM = 1.14 eV), in 24.09 (at %) as NbO for BE = 203.24 eV (FWHM = 1.32 eV), in 8.76 (at %) as $NbO_2$ for BE = 205.66 eV (FWHM =

2.02 eV) and in 5.66 (at %) as $Nb^0$ for BE = 201.77 eV (FWHM = 0.79 eV) [26-27]. There is also a compound formed with niobium with the BE of 204.28 eV, which is the location of NbC or $NbN_x$ [26]. But it is impossible to determine whether it is NbC or a $NbN_x$ only based on the XPS spectra of Nb 3d.

Oxygen in the surface layer formed after nitrogen doping treatment with the 2/6 recipe without subsequent EP removal mainly occurs in 59.61 (at %) as $Nb_2O_5$ for BE = 530.08 eV (FWHM = 1.18 eV) [28], in 34.30 (at %) as NbO for BE = 531.11 eV (FWHM = 2.51 eV) [29] and in 5.52 (at %) as a compound possibly being $CF_3SSOOCH_3$ for BE = 532.48 eV (FWHM = 1.53 eV) [30]. There is also a trace of oxygen occurring in 0.58 (at %) as $H_2O$ for BE = 533.53 eV (FWHM = 0.81 eV) [31].

Carbon in the surface layer formed after nitrogen doping treatment with the 2/6 recipe without subsequent EP removal mainly occurs in 75.51 (at %) as hydrocarbon for BE = 284.83 eV (FWHM = 1.33 eV) [32]. Carbon also forms compounds with other elements on the surface of the sample, for example, in 10.04 (at %) as the formation of $CH_3CN$ or $(CH_3CH_2)_2O$ for BE = 286.36 eV (FWHM = 1.40 eV) [33], in 9.09 (at %) as the formation of $C_6F_6$ or $C_6H_4F_2$ for BE = 288.33 eV (FWHM = 1.78 eV) [34] and in 5.35 (at %) as the formation of NbC or TiC for BE = 282.25 eV (FWHM = 1.06 eV) [35].

The signal peak of nitrogen has good symmetry, which means the chemical distribution of N is relatively simple. Compared with the standard energy spectrum, nitrogen in the surface layer formed after nitrogen doping treatment with the 2/6 recipe without subsequent EP removal may mainly possibly occur as $CH_3CN$ for BE = 399.84eV (FWHM = 1.69 eV) [36]. No nitride [37] was detected in the surface layer of the sample directly after nitrogen doping treatment with the 2/6 recipe.

According to the chemical distribution of carbon, nitrogen and niobium, it can be seen that the compound formed with niobium for the BE of 204.28 eV (FWHM = 0.60 eV) is NbC but not $NbN_x$ and its content is 4.56 (at %). The elemental composition and chemical states on the surfaces of all niobium samples with different treatments were basically the same. No nitrides were found on any surface of all the samples.

*3.2 In the penetration depth*

We selected five samples from the experiments described earlier for the GCIB etching, which includes the ND-2[nd]-0μm sample, the ND-2[nd]-7μm sample, the ND-2[nd]-21μm sample, the HT-2[nd]-20μm sample and the noNDnoHT sample. High resolution spectra for Nb 3d (a), N 1s (b), O 1s (c) and C 1s (d) on the surface of ND-2[nd]-0μm sample after different amounts of GCIB etching are presented in Fig. 4 for reference.

For C element, after GCIB etching, within the depth below 30nm from the surface layer, the interior of the sample is substantially free of hydrocarbon and C mainly exists in the form of NbC or TiC. For O element, the main distribution form is $Nb_2O_5$ both before and after GCIB etching. However, in the depth range below the surface layer of 30nm, due to the reduction or absence of C, F, S, the other compositions of O is not detected. The GCIB etching depth exceeds the depth of niobium oxide distribution, so niobium mainly occurs as $Nb^0$ after GCIB etching. But the amount of nitrogen diffused into niobium is minimal. So after the GCIB etching depth of more than 30nm, no effective signal peaks for N were detected during the collection time of 5 minutes.

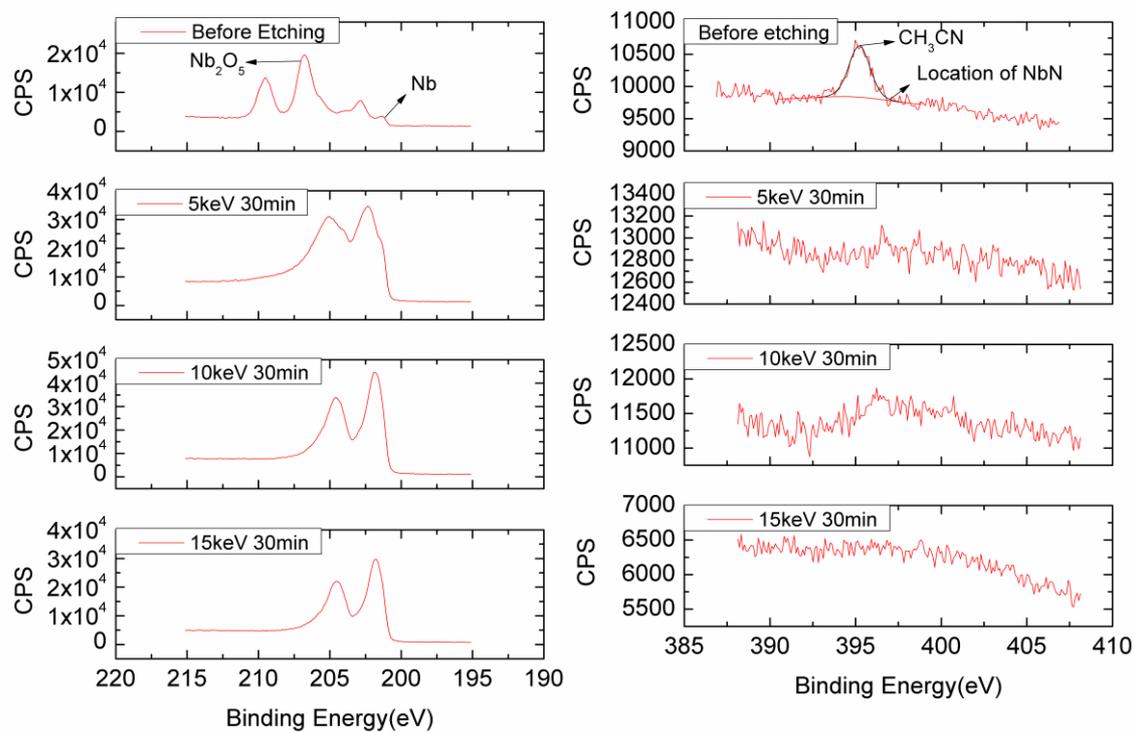

(a)                                                          (b)

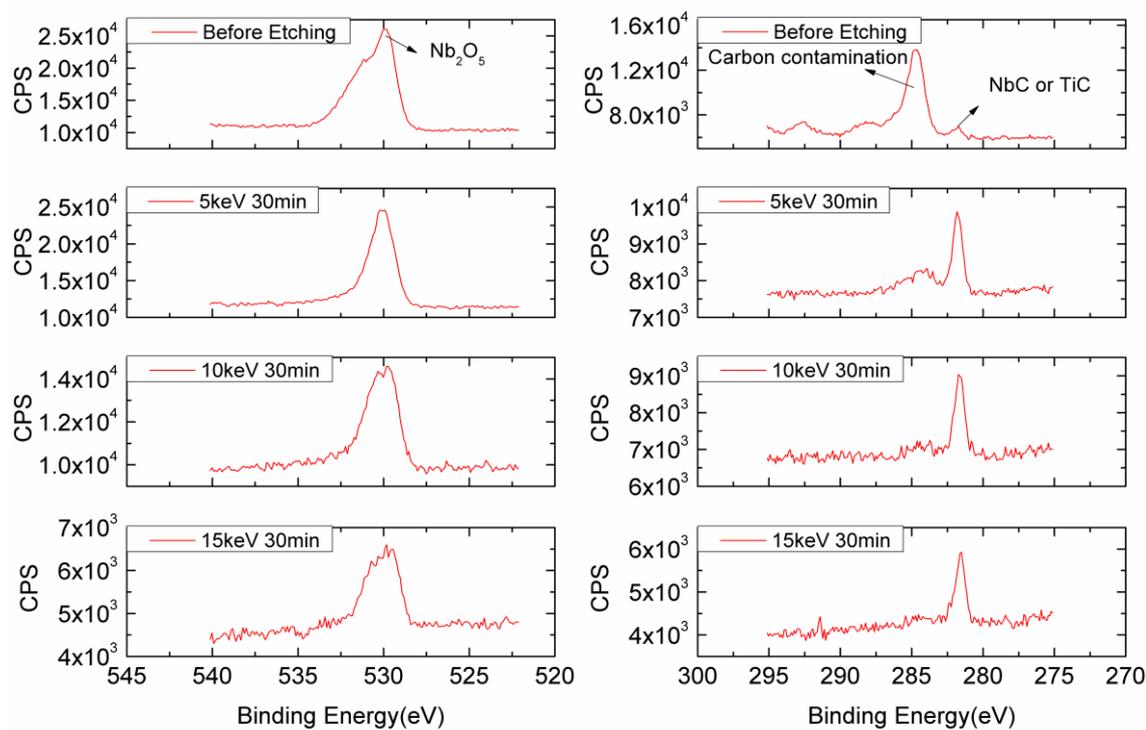

(c)                                                          (d)

**Fig. 4.** High resolution XPS spectra for Nb 3d (a), N 1s (b), O 1s (c) and C 1s (d) on the surface of ND-2$^{nd}$-0μm sample after different amounts of GCIB etching.

XPS test results of other samples after GCIB etching is similar to that of the ND-2$^{nd}$-0μm sample.

## 4. SEM and XRD results

Scanning electron microscopy (SEM) experiments were performed on ND-1$^{st}$-0μm sample and ND-2$^{nd}$-0μm sample to reveal surface features of niobium directly after nitrogen doping treatment with the 2/6 recipe. As it can be seen in Fig. 5, no star-shaped features [16] were found on the surface of the two doped samples with the 2/6 recipe.

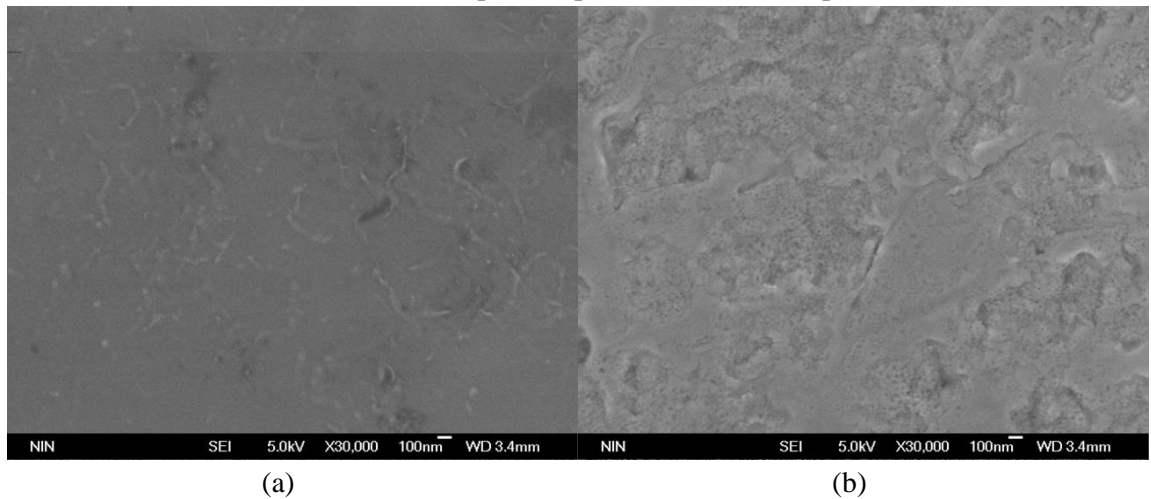

(a) (b)

**Fig. 5.** SEM images of ND-1$^{st}$-0μm sample (a) and ND-2$^{nd}$-0μm sample (b).

The X-ray diffraction analysis was performed to detect the surface compositions to verify the experimental results of XPS. X-ray powder diffraction patterns of the ND-2$^{nd}$-0μm sample and the ND-2$^{nd}$-1μm sample are shown in Fig. 6. Peaks heights have been normalized with respect to the strongest one and the graphs abscissa range goes from 10 to 100$^0$. Due to the much deeper XRD detection depth than XPS, almost all the peaks have been associated to the Nb phase. Small peaks of Nb$_2$O$_5$, NbO and carbon compounds can be noticed on both sample surfaces. Nb$_2$C trace was found on the surface of ND-2$^{nd}$-0μm sample while no signals of nitrides were detected on both samples surfaces with the 2/6 recipe, which is in confirmation with the XPS results.

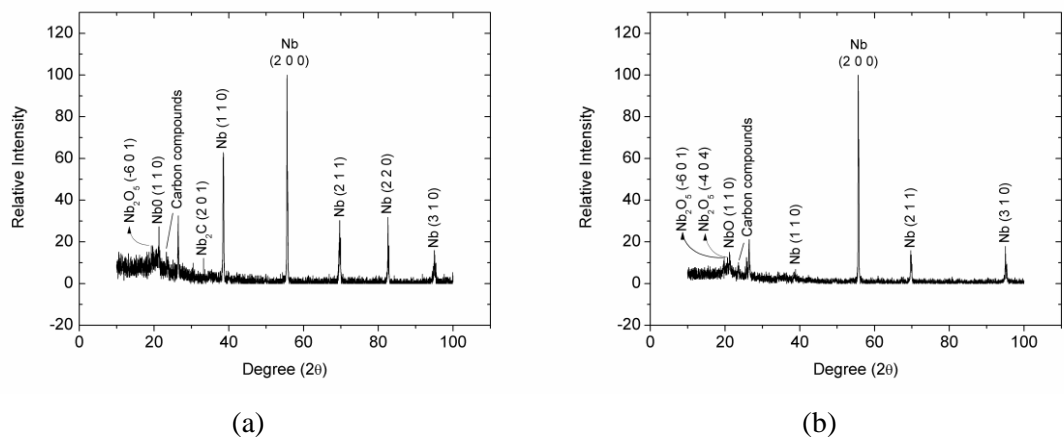

(a) (b)

**Fig. 6.** X-ray powder diffraction patterns of ND-2$^{nd}$-0μm sample (a) and ND-2$^{nd}$-1μm sample

(b).

## 5. Discussion

The surface resistance determines the Q value of the SRF cavities directly, therefore affects the cryogenic load of the superconducting accelerator. The subtle material details of the niobium surface largely influence the surface resistance of the SRF cavities. It is commonly known that different surface treatment techniques result in a variety of possible surface compositions.

Fundamental understanding of the nitrogen doping mechanism has been being carried out extensively, but yet remains unclear. Formation of niobium nitride phases was observed in Nb samples annealed in nitrogen atmosphere at $800^0$C for 10 minutes by XRD [38] and in Nb samples annealed in nitrogen atmosphere at $800^0$C for 20 minutes by TEM [39]. As mentioned above, vertical test results of N-doped SRF cavities with varying doping parameters showed that the 2/6 recipe proposed by FNAL is the best nitrogen doping recipe. Thus in this paper, we focused on the 2/6 recipe and investigated the surface composition of N-doped niobium used for the fabrication of SRF cavities by using XPS analysis.

The surface chemical composition of Nb samples both on the surfaces and in the penetration depth with different post treatments has been studied by XPS. Combined with the TOF-SIMS measurements of the nitrogen doping niobium samples [18], the XPS detection depth is about 5 - 7 nm, which covers the oxide and the oxide/metal interface. Within the detection depth of 5 - 7 nm and within the detection area of about 500 μm × 500 μm, analysis of the composition and chemical states of elements on the sample surfaces was carried out from the narrow scan spectra with higher energy resolution of 0.05 eV. Niobium on the surfaces of all samples occurs mainly as $Nb_2O_5$, NbO, $NbO_2$ and NbC. Oxygen on the surfaces of all samples occurs mainly as $Nb_2O_5$ and also as some compounds formed together with other elements in the surface layer. Carbon on the surfaces of all samples occurs mainly as hydrocarbon which can be considered as contamination. No signals of poorly superconducting nitrides $NbN_x$ was found on the surface of any doped sample with the 2/6 recipe. Nitrogen in the surface layers of all Nb samples, including both the doped and un-doped samples, occurs as a compound for the same BE of about 399.84 eV. According to the standard energy spectrum, this compound may be $CH_3CN$. No nitrides were detected from both the SEM images and XRD analyses of the 2/6 doped Nb samples without subsequent EP removal. It means that the poor quality factor of 2/6 N-doped Nb SRF cavities without subsequent EP material removal may be not due to the formation of poorly superconducting nitrides. Since only $NbC_x$ signal was detected, niobium carbides may be a possible cause of the 2/6 N-doped cavity's extremely low quality factor before EP treatment.

After GCIB etching, within the depth below 30nm from the surface layer, the content of N is lower than the XPS detection accuracy. It means that the doped nitrogen of the 2/6 recipe is trace and special attention is needed to be paid to the control of the nitrogen doping conditions.

## 6. Conclusion

The elemental composition and chemical states both on the surfaces and in the penetration depth range of niobium samples with different treatments have been carried out by using XPS. Niobium on the surfaces of all samples occurs mainly as $Nb_2O_5$, $NbO$, $NbO_2$ and $NbC$. Oxygen on the surfaces of all samples occurs mainly as $Nb_2O_5$ and also as some compounds formed together with other elements in the surface layer. Carbon on the surfaces of all samples occurs mainly as hydrocarbon. Nitrogen on the surfaces of all samples may mainly occur as $CH_3CN$. No signals of poorly superconducting nitrides $NbN_x$ was found on the surface of any doped sample with the 2/6 recipe, which was confirmed by the SEM and XRD results. It means that the poor quality factor of 2/6 N-doped Nb SRF cavities without subsequent EP material removal is possibly not due to the formation of poorly superconducting nitrides.

After GCIB etching, within the depth below 30nm from the surface layer, the interior of the samples is substantially free of hydrocarbons and C mainly exists in the form of NbC or TiC. For O element, the main distribution form is $Nb_2O_5$ both before and after GCIB etching. For Nb element, niobium mainly occurs as $Nb^0$ after GCIB etching. After the GCIB etching depth of more than 30nm, no effective signal peaks for N were detected during the collection time of 5 minutes, which means that the doped nitrogen of the 2/6 recipe is trace and special attention is needed to be paid to the control of the nitrogen doping conditions.

**Acknowledgments**

The authors are grateful to Lin Lin (Peking University) for his help in EP. The authors would also acknowledge Dr. J. L. Xie (Peking University) and Dr. Z. J. Zhao (Institute of Chemistry, Chinese Academy of Science) for the help in XPS measurements. This work is supported by Major Research Plan of National Natural Science Foundation of China (91426303) and National Postdoctoral Program for Innovative Talents (BX201700257).